\documentstyle[12pt]{article}
\begin{document}
\def\beq{\begin{equation}}
\def\eeq{\end{equation}}
\def\bea{\begin{eqnarray}}
\def\eea{\end{eqnarray}}
\def\ve{\vert}
\def\vel{\left|}
\def\ver{\right|}
\def\nnb{\nonumber}
\def\ga{\left(}
\def\dr{\right)}
\def\aga{\left\{}
\def\adr{\right\}}
\def\rar{\rightarrow}
\def\nnb{\nonumber}
\def\la{\langle}
\def\ra{\rangle}
\def\lla{\left<}
\def\rra{\right>}
\def\ba{\begin{array}}
\def\ea{\end{array}}
\def\tep{$B \rar K \ell^+ \ell^-$}
\def\tepm{$B \rar K \mu^+ \mu^-$}
\def\tept{$B \rar K \tau^+ \tau^-$}
\def\ds{\displaystyle}



\def\bos{\lower 0.5cm\hbox{{\vrule width 0pt height 1.2cm}}}
\def\boss{\lower 0.35cm\hbox{{\vrule width 0pt height 1.cm}}}
\def\aaa{\lower 0.cm\hbox{{\vrule width 0pt height .7cm}}}
\def\dol{\lower 0.4cm\hbox{{\vrule width 0pt height .5cm}}}


\title{ {\Large {\bf 
Exclusive $\Lambda_b \rar \Lambda \ell^+ \ell^-$ decay 
in two Higgs doublet model} } }

\author{\vspace{1cm}\\
{\small T. M. Aliev \thanks
{e-mail: taliev@metu.edu.tr}\,\,,
M. Savc{\i} \thanks
{e-mail: savci@metu.edu.tr}} \\
{\small Physics Department, Middle East Technical University} \\
{\small 06531 Ankara, Turkey} }
\date{}

\begin{titlepage}
\maketitle
\thispagestyle{empty}

\begin{abstract}
\baselineskip  0.7cm
Rare $\Lambda_b \rar \Lambda \ell^+ \ell^-$ decay is investigated 
in framework of general  two Higgs doublet model, in which a new source of
CP violation exists (model III). The polarization parameter,
CP asymmetry and decay width are calculated. It is shown that CP asymmetry
is a very sensitive tool for establishing model III.
\end{abstract}


\end{titlepage}

\section{Introduction}
Rare decays, induced by flavor--changing neutral current (FCNC) 
$b \rar s(d)$ transitions, provide testing grounds for the standard 
model (SM) at loop level and can give 
valuable information about the Cabibbo--Kobayashi--Maskawa (CKM) 
matrix elements $V_{td},~V_{ts}$, $V_{tb}$, etc. In addition, the study 
of rare decays can pave the way for establishing new physics beyond SM, 
such as two Higgs doublet model (2HDM), minimal supersymmetric extension 
of the SM (MSSM). Most important of all, study of the 
$b \rar s(d) \ell^+ \ell^-$ decays is expected to be one of the most 
reliable quantitative tests of FCNC. This transition has been extensively 
investigated in framework of the SM, 2HDM and MSSM \cite{R1}--\cite{R16}.  

The matrix element of the 
$b \rar s \ell^+ \ell^-$ contains terms describing the virtual effects
induced by $t\bar t$,  $c\bar c$  and $u\bar u$ loops which are proportional
to $V_{tb}\,V_{ts}^*$, $V_{cb}\,V_{cs}^*$ and $V_{ub}\,V_{us}^*$,
respectively. Using unitarity of the CKM matrix and neglecting
$V_{ub}\,V_{us}^*$ in comparison to $V_{tb}\,V_{ts}^*$ and
$V_{cb}\,V_{cs}^*$, it is obvious that the matrix element for the 
$b \rar s \ell^+ \ell^-$ involves only one independent CKM factor 
$V_{tb}\,V_{ts}^*$ so that CP--violation in this channel is strongly
suppressed in the SM. 

The present work is to devoted studying the 
exclusive $\Lambda_b \rar \Lambda \ell^+ \ell^-$ 
decay, which at quark level is described by $b \rar s \ell^+ \ell^-$
transition, in context of the general two Higgs doublet
model in which a new source for CP violation exists. 

2HDM model is one of the simplest extension of the SM, which contains 
two complex Higgs doublets, while the SM contains only one. In general, 
in 2HDM the flavor changing neutral currents (FCNC) that appear at tree 
level, are avoided by imposing an {\it ad hoc} discrete symmetry \cite{R17}. 
One possible solution to avoid these unwanted FCNC at tree level is that
all fermions couple to only one of the above--mentioned Higgs doublets
(model I). The other possibility is the coupling of the up and down quarks
to the first and second Higgs doublets, with the vacuum expectation values 
$v_2$ and $v_1$, respectively (model II). Model II is more attractive since
its Higgs sector coincides with the ones in the supersymmetric model.
The strength of couplings of fermions with Higgs fields depends on
$\tan \beta=v_2/v_1$, which is the free parameter of the model. The new
experimental results of CLEO and ALEPH Collaborations \cite{R18,R19} 
on the branching ratio
$b \rar s \gamma$ decay impose strict restrictions on the charged Higgs
boson mass and $\tan \beta$. Recently, the lower bound on these parameters
were determined from the analysis of the $b \rar s \gamma$ decay, including            
NLO QCD corrections \cite{R20,R21}.
The phenomenological consequence of a more general model in 2HDM, namely,
model III, without discrete symmetry has been investigated in 
\cite{R22}--\cite{R24}. In this model FCNC appears 
naturally at tree level. However, the FCNC's involving the first two
generations are highly suppressed, as is observed in the low energy
experiments, and those involving the third generation is not as severely
suppressed as the first two generations, which are restricted by the
existing experimental results. 

In this work we assume that all tree level FCNC couplings are negligible.
However even with this assumption, the couplings of 
fermions to Higgs bosons may have a complex phase $e^{i\phi}$. 
In other words, in this model there exists a new source of CP violation that
is absent in the SM, model I and model II.
The effects of such an extra phase in the $b \rar s \gamma$ and 
$b \rar d \ell^+ \ell^-$, $B \rar \pi \ell^+ \ell^-$ and
$B \rar \rho \ell^+ \ell^-$ decays were discussed
in \cite{R25,R26} and \cite{R27}, respectively. 
The constraints on the phase angle $\phi$ in the product 
$\lambda_{tt} \lambda_{bb}$ of Higgs--fermion coupling (see below) 
imposed by the neutron electric dipole moment, 
$B^0 - \bar B^0$ mixing. $\rho \,_0$ parameter and $R_b$ is discussed in 
\cite{R26}.        

The paper is
organized as follows: In Section 2 we present the necessary theoretical
framework and the branching ratio, CP--violating effects in the partial
widths for the above--mentioned 
exclusive decay channels are studied. Section 3 is devoted to
the numerical analysis and concluding remarks.       
  
\section{Theoretical calculations for the 
$\Lambda_b \rar \Lambda \ell^+ \ell^-$ decay}

Before presenting the theoretical results for
$\Lambda_b \rar \Lambda \ell^+ \ell^-$ decay, let us remember
the main essential points of the general Higgs doublet model (model III). 
Without loss of generality we can
work in a basis such that only the first doublet generates all the fermion and
gauge boson masses, whose vacuum expectation values are
\bea
\left< \phi_1 \right> = \left( \begin{array}{c}
0 \\ \\
\displaystyle{\frac{v}{\sqrt{2}}}	
\end{array} \right)~~~~~,~ \left< \phi_2 \right>=0~. \nnb
\eea
In this basis the first doublet $\phi_1$ is the same as in the SM, and all
new Higgs bosons result from the second doublet $\phi_2$, which can be
written in the following form
\bea
\phi_1 = \frac{1}{\sqrt{2}} \left( \begin{array}{c}
\sqrt{2}\, G^+ \\ \\
v + \chi_1^0 + i G^0 
\end{array} \right)~~~~~,
~ \phi_2  = \frac{1}{\sqrt{2}} \left( \begin{array}{c}
\sqrt{2}\, H^+ \\ \\
\chi_2^0 + i A^0 
\end{array} \right)~, \nnb
\eea
where $G^+$ and $G^0$ are the Goldstone bosons. The neutral $\chi_1^0$ and
$\chi_2^0$ are not the physical mass eigenstate, but their linear
combinations give the neutral $H^0$ and $h^0$ Higgs bosons:
\bea
\chi_1^0 = H^0 \cos \alpha - h^0 \sin \alpha~, \nnb \\
\chi_2^0 = H^0 \sin \alpha + h^0 \cos \alpha~. \nnb
\eea
The general Yukawa Lagrangian can be written as 
\bea
{\cal L}_Y = \eta_{ij}^U \bar Q_{iL} \tilde \phi_1 U_{jR} +
\eta_{ij}^{\cal D} \bar Q_{iL} \phi_1 {\cal D}_{jR}
+ \xi_{ij}^U \bar Q_{iL} \tilde \phi_2  U_{jR} +
\xi_{ij}^{\cal D} \bar Q_{iL} \phi_2  {\cal D}_{jR} + h.c.~,
\eea
where $i$, $j$ are the generation indices, $\tilde \phi= i \sigma_2 \phi$,
$\eta_{ij}^{U,{\cal D}}$ and $\xi_{ij}^{U,{\cal D}}$, in general, are the
non--diagonal coupling matrices, $L=(1-\gamma_5)/2$ and $R=(1+\gamma_5)/2$ are
the left-- and right--handed projection operators. In Eq. (1) all states
are weak states, that can be transformed to the mass eigenstates by
rotation. In mass eigenstates the Yukawa Lagrangian is 
\bea
{\cal L}_Y = - H^+ \bar U \left[ V_{CKM} \hat \xi^{\cal D} R - 
\hat \xi^{U^+} V_{CKM} L \right] {\cal D}~,
\eea
where $U({\cal D})$ represents the mass eigenstates of $u,~c,~t~(d,~s,~b)$
quarks. In this work, we will use a simple ansatz for
$\hat \xi^{U^+,{\cal D}}$ \cite{R22},
\bea
\hat \xi^{U^+,{\cal D}} = \lambda_{ij} 
\frac{g \sqrt{m_i m_j}}{\sqrt{2} m_W}~,
\eea
assume that $\lambda_{ij}$ is complex, i.e., 
$\lambda_{ij} = \vel \lambda_{ij} \ver e^{i\phi}$. For simplicity we
choose $\xi^{U,{\cal D}}$ to be diagonal to suppress all tree level FCNC
couplings, and as a result $\lambda_{ij}$ are also diagonal but remain
complex. Note that the results for model I and model II can be obtained from
model III by the following substitutions:
\bea 
&&\lambda_{tt} = \cot \beta ~~~~~~~ \lambda_{bb} = - \, \cot \beta ~~\mbox{\rm
for model I}~, \nnb \\
&&\lambda_{tt} = \cot \beta ~~~~~~~ \lambda_{bb} = + \, \tan  \beta ~~\mbox{\rm   
for model II}~,
\eea
and $\phi = 0$.

After this brief introduction about the general Higgs doublet model,
let us return our attention to the $b \rar s \ell^+ \ell^-$
decay. The powerful framework into which the 
perturbative QCD corrections to the physical decay amplitude incorporated 
in a systematic way, is the effective Hamiltonian method. 
In this approach, the heavy degrees of freedom, 
$t$ quark, $W^\pm,~H^\pm,~h^0,~H^0$ are integrated out. 
The procedure is to match the
full theory with the effective theory at high scale $\mu = m_W$, and then
calculate the Wilson coefficients at lower $\mu \sim {\cal O}(m_b)$ using the 
renormalization group equations. In our calculations we choose the higher
scale as $\mu = m_W$, since the charged Higgs boson is heavy enough
($m_{H^\pm} \ge 210 ~GeV$ see \cite{R20}) to neglect the evolution from 
$m_{H^\pm}$ to $m_W$. 

In this work the charged Higgs boson contributions are taken into
account and the neutral Higgs boson exchange diagram contributions are
neglected since Higgs--fermion interaction is proportional to the lepton
mass. The charged Higgs boson exchange diagrams do not
produce new operators and the operator basis is the same as the one used 
for the $b \rar s \ell^+ \ell^-$ decay in the SM.
Therefore in model III, 
the charged Higgs boson contributions to leading order change only the value
of the Wilson coefficients at $m_W$ scale, i.e.,
\bea
C_7^{2HDM}(m_W) &=& C_7^{SM}(m_W) + C_7^{H^\pm}(m_W) \nnb \\
C_9^{2HDM}(m_W) &=& C_9^{SM}(m_W) + C_9^{H^\pm}(m_W) \nnb \\
C_{10}^{2HDM}(m_W) &=& C_{10}^{SM}(m_W) + C_{10}^{H^\pm}(m_W)~. \nnb
\eea
The coefficients $C_i^{2HDM} (m_W)$ to the leading order are given by
\bea
C_7^{2HDM}(m_W) &=& 
x \, \frac{(7-5 x - 8 x^2)}{24 (x-1)^3} + 
\frac{x^2 (3 x - 2)}{4 (x-1)^4} \, \ln x \nnb \\
&+& \vel \lambda_{tt} \ver^2 \Bigg[ \frac{y(7-5 y - 8 y^2)}
{72 (y-1)^3} + \frac{y^2 ( 3 y - 2)}{12 (y-1)^4} \, \ln y \Bigg] \nnb \\
&+& \lambda_{tt} \lambda_{bb} \Bigg[ \frac{y(3-5 y)}{12 (y-1)^2} +
\frac{y (3 y - 2)}{6 (y-1)^3} \, \ln y \Bigg]~, \\ \nnb \\ \nnb \\
C_9^{2HDM}(m_W) &=& - \frac{1}{\sin^2 \theta_W} \, B(m_W) + 
\frac{1 - 4 \sin^2 \theta_W}{\sin^2 \theta_W} \, C(m_W) \nnb \\
&+& \frac{-19 x^3 + 25 x^2}{36 (x-1)^3} +
\frac{-3 x^4 + 30 x^3 - 54 x^2 + 32 x -8}{18 (x-1)^4} \, \ln x 
+ \frac{4}{9} \nnb \\
&+& \vel \lambda_{tt} \ver^2 \Bigg\{ 
\frac{1 - 4 \sin^2 \theta_W}{\sin^2 \theta_W} \, \frac{x y}{8} \Bigg[ 
\frac{1}{y-1} - \frac{1}{(y-1)^2} \, \ln y \Bigg]\nnb \\
&-& y \Bigg[ \frac{47 y^2 - 79 y + 38}{108 (y-1)^3}
-\frac{3 y^3 - 6 y^3 + 4}{18 (y-1)^4} \, \ln y \Bigg] \Bigg\}~,
\\ \nnb \\ \nnb \\
C_{10}^{2HDM}(m_W) &=& \frac{1}{\sin^2 \theta_W} \Big[ B(m_W) - 
C(m_W) \Big] \nnb \\
&+& \vel \lambda_{tt} \ver^2 \frac{1}{\sin^2 \theta_W} \,\frac{x y}{8} 
\Bigg[ - \frac{1}{y-1} + \frac{1}{(y-1)^2} \, \ln y \Bigg]~, 
\eea
where
\bea
B(x) &=& - \frac{x}{4 (x-1)} + \frac{x}{4 (x-1)^2} \, \ln x ~, \nnb \\
C(x) &=& - \frac{x}{4} \Bigg[ \frac{x-6}{3 (x-1)} + 
\frac{3 x +2 }{2 (x-1)^2} \ln x \Bigg]~,\nnb \\
x &=& \frac{m_t^2}{m_W^2} ~, \nnb \\
y &=& \frac{m_t^2}{m_{H^\pm}^2}~.
\eea
and $\sin^2\theta_W = 0.23$ is the Weinberg angle. 
It follows from Eqs. (5--8)
that among all the Wilson coefficients, only $C_7$ involves the new phase
angle $\phi$.

The effective Hamiltonian for the $b \rar s \ell^+ \ell^-$ decay is 
[28--31]
\bea
{\cal H} = -4 \frac{G_F}{2\sqrt 2} V_{tb} V^*_{ts}
\sum_{i=0}^{10} C_i(\mu) O_i(\mu)~, \nnb
\eea
where 
$C_i$ are the Wilson coefficients.
The explicit form of all operators $O_i$ can be found in [28--31].

The evolution of the Wilson coefficients from the higher scale 
$\mu = m_W$ down to the low energy scale $\mu = m_b$ is described by the
renormalization group equation
\bea
\mu \frac{d}{d\mu}C_i^{eff(\mu)} = 
C_i^{eff}(\mu) \gamma_\mu^{eff}(\mu)~, \nnb
\eea
where $\gamma$ is the anomalous dimension matrix. 
The coefficient $C_7^{eff}(\mu)$ at the scale 
$\mu=m_b$ in next to leading order (NLO) 
is calculated in \cite{R20,R21}:
\bea
C_7^{eff}(m_b) = C_7^0(m_b) + \frac{\alpha_s(m_b)}{4 \pi} 
C_7^{1,eff}(m_b)~,\nnb
\eea
where $C_7^0(m_b)$ is the leading order (LO) term and $C_7^{1,eff}(m_b)$
describes the NLO terms, whose explicit forms can be found in 
\cite{R20}. In our case,
the expressions for these coefficients can be obtained from the results of 
\cite{R21} by making the following replacements:
\bea
\vel Y \ver^2  \rar \vel \lambda_{tt} \ver^2 ~~~~~ \mbox{\rm and} ~~~~~ 
X Y^*  \rar \vel \lambda_{tt} \lambda_{bb} \ver e^{i\phi}~. \nnb
\eea
In the SM, the QCD corrected Wilson coefficient $C_9(m_b)$, which
enters to the decay amplitude up to the next leading order has been
calculated in [28--31]. The Wilson coefficient $C_{10}$ is not modified as
we move from $\mu=m_W$ to $\mu=m_b$ scale, 
i.e., $C_{10}(m_b) \equiv C_{10}^{2HDM}(m_W)$.  
As we have already noted, in model III
there does not appear any new operator other than
those that exist in the SM, therefore it is enough to make the
replacement $C_9^{SM}(m_W) \rar C_{9}^{2HDM}(m_W)$ in [28--31], in order
to calculate $C_{9}^{2HDM}$ at $m_b$ scale. Hence, including the NLO
QCD corrections, $C_9(m_b)$ can be written as:
\bea
\lefteqn{
C_9(\mu) = C_9^{2HDM}(\mu) 
\left[1 + \frac{\alpha_s(\mu)}{\pi} \omega (\hat s) \right]} \nnb \\
&&+ \, g(\hat m_c,\hat s) \Big[ 3 C_1(\mu) + C_2(\mu) + 3 C_3(\mu) + C_4(\mu)
+ 3 C_5(\mu) + C_6(\mu) \Big] \nnb \\
&&- \frac{1}{2} g(0,\hat s) \ga C_3(\mu) + 3 C_4(\mu) \dr
-\,  \frac{1}{2} g \ga 1, \hat s\dr
\ga 4 C_3 + 4 C_4 + 3 C_5 + C_6 \dr \nnb \\
&&- \frac{1}{2} g \ga 0, \hat s\dr \ga C_3 + 3 C_4 \dr
+\, \frac{2}{9} \ga 3 C_3 + C_4 + 3 C_5 + C_6 \dr~, 
\eea
where $\hat m_c = m_c/m_b~, ~\hat s = p^2/m_b^2$, and
\bea
\lefteqn{
\omega \ga \hat s \dr = - \frac{2}{9} \pi^2 - 
\frac{4}{3} Li_2  \ga \hat s \dr - \frac{2}{3} \ln \ga \hat s\dr 
\,\ln \ga 1 -\hat s \dr} \nnb \\ 
&&- \,\frac{5 + 4 \hat s}{3 \ga 1 + 2 \hat s \dr} \ln \ga 1 -\hat s \dr
-\frac{2 \hat s \ga 1 + \hat s \dr \ga 1 - 2 \hat s \dr}
{3 \ga 1 - \hat s \dr^2 \ga 1 + 2 \hat s \dr} \, \ln \ga \hat s\dr 
+ \frac{5 + 9 \hat s - 6 {\hat s}^2}
{3 \ga 1 - \hat s \dr \ga 1 + 2 \hat s \dr}~
\eea
represents the ${\cal O}\ga \alpha_s \dr$ correction from the one gluon
exchange in the matrix element of $O_9$, while the function
$g \ga \hat m_c, \hat s \dr$ arises from one loop contributions of the
four--quark operators $O_1$--$O_6$, whose form is
\bea 
g \ga \hat m_c, \hat s \dr &=& - \frac{8}{9}  \ln \ga \hat m_i\dr
+ \frac{8}{27} + \frac{4}{9} y_i \nnb \\
&&- \frac{2}{9} \ga 2 + y_i \dr
 \sqrt{\vel 1-y_i \ver} \Bigg\{ \Theta \ga 1 - y_i \dr 
\Bigg( \ln \frac{1+\sqrt{\vel 1-y_i \ver}}{1-\sqrt{\vel 1-y_i \ver}}
- i \, \pi \Bigg) \nnb \\
&&+ \Theta \ga y_i -1 \dr 2 \arctan \frac{1}{\sqrt{y_i - 1}} \Bigg\}~,
\eea
where $y_i = 4 {\hat m_i}^2/{\hat p}^2$.

The Wilson coefficients $C_9$ receives also long distance contributions,
which have their origin in the real $c\bar c$ 
intermediate states, i.e., $J/\psi$, $\psi^\prime$,
$\cdots$. The $J/\psi$ family is  
introduced by the Breit--Wigner distribution for the resonances
through the replacement ([4--7,32])
\bea
g \ga \hat m_c, \hat s \dr \rar g \ga \hat m_c, \hat s \dr -
\frac{3\pi}{\alpha^2_{em}}
\, \kappa \sum_{V_i=J/\psi_i,\psi^\prime,\cdots} 
\frac{m_{V_i} \Gamma(V_i \rar \ell^+ \ell^-)}
{(p^2 - m_{V_i}^2) + i m_{V_i} \Gamma_{V_i}}~,
\eea
where the phenomenological parameter $\kappa =2.3$ is chosen in order to
reproduce correctly the experimental value of the branching ratio
(see for example \cite{R15})
\bea
{\cal B}(B \rar J/\psi X \rar X \ell^+ \ell^-)={\cal B}(B \rar J/\psi X)   
\,{\cal B}(J/\psi \rar X \ell^+ \ell^-)~. \nnb
\eea

The effective short--distance Hamiltonian for $b \rar s\ell^+ \ell^-$
decay [28--31] leads to the QCD corrected matrix element (when the $s$
quark mass is neglected)
\bea
{\cal M} &=& \frac{G_F \alpha_{em}}{2\sqrt 2 \pi} V_{ts} V^*_{tb} \Bigg\{
C_9^{eff}(m_b) \bar s \gamma_\mu (1- \gamma_5) b \, \bar \ell \gamma^\mu \ell + 
C_{10}(m_b) \bar s \gamma_\mu (1- \gamma_5) b \, \bar \ell 
\gamma^\mu \gamma_5 \ell \nnb \\
&-& 2 C_7^{eff}(m_b)\frac{m_b}{p^2}\bar s i \sigma_{\mu \nu}p^\nu (1+\gamma_5)  b  \,
\bar \ell \gamma^\mu \ell\Bigg\}~,
\eea
where $p^2$  is the invariant dilepton mass.

After obtaining the matrix element for $b \rar s \ell^+ \ell^-$ transition,
our next task is, starting from this matrix element,  
to calculate the matrix element of the 
$\Lambda_b \rar \Lambda \ell^+ \ell^-$ decay. It follows from the matrix
element of the $b \rar s \ell^+ \ell^-$ that, the matrix elements 
$\la \Lambda \ve \bar s \gamma_\mu (1 - \gamma_5 ) b \ve \Lambda_b \ra$ and  
$\la \Lambda \ve \bar s i\, \sigma_{\mu\nu}p_\nu (1 + \gamma_5 ) b
\ve \Lambda_b \ra$ have to be calculated in order in order to be able to
calculate the matrix element of the exclusive
$\Lambda_b \rar \Lambda \ell^+ \ell^-$ decay. A lot of form factors are
required for a description of this decay. However when 
the heavy quark effective theory (HQET) has been used, the heavy quark 
symmetry reduces the number of independent form factors 
for the baryonic transition $\Lambda_Q \rar$ light spin--1/2 baryon,
only to two 
$(F_1$ and $F_2)$, irrelevant
to the Dirac structure of the relevant operators (for more details see
\cite{R33}) 
\bea
\lla \Lambda(p,s) \vel \bar s \Gamma b \ver \Lambda_b(v,s^\prime) \rra =
\bar u_\Lambda(p,s) \Big\{ F_1 (pv) + \not\! v F_2(pv) \Big\}
\Gamma u_{\Lambda_b} (v,s^\prime)~,
\eea
where $v$ is the four--velocity of $\Lambda_b$, $\Gamma$ is an arbitrary 
Dirac structure (in our case $\Gamma = \gamma_\mu (1- \gamma_5)$ and 
$i \sigma_{\mu \nu}p^\nu (1+\gamma_5)$).
The form factors $F_1$ and $F_2$ for the 
$\Lambda_b \rar \Lambda \ell^+ \ell^-$ decay are calculated in framework 
of the QCD sum rules approach in \cite{R34}.   
So the matrix element of the $\Lambda_b \rar \Lambda \ell^+ \ell^-$ decay
takes the following form:
\bea
{\cal M} &=& \frac{G_F \alpha_{em}}{2\sqrt 2 \pi} V_{ts} V^*_{tb} \Bigg\{
\bar u_\Lambda(p,s) \Big[F_1 + F_2 \not\! v \Big] \gamma_\mu(1 - \gamma_5 ) 
u_{\Lambda_b}(v,s^\prime) \Big[C_9^{eff} \bar \ell \gamma_\mu \ell + 
C_{10}\bar \ell \gamma_\mu \gamma_5 \ell \Big] \nnb \\
&-& C_7^{eff}\frac{m_b}{p^2} \bar u_\Lambda(p,s) \Big[F_1 + F_2 \not\! v \Big] 
i\, \sigma_{\mu\nu}p_\nu (1 + \gamma_5 ) u_{\Lambda_b}(v,s^\prime)
\bar \ell \gamma_\mu \ell \Bigg\}~.
\eea
Using Eq. (15) and summing over polarization of the final
leptons and averaging over polarization of the initial $\Lambda_b$, we get
the following result for the double differential decay rate (the
masses of the final leptons are neglected and all calculations are performed
in the rest frame of the $\Lambda_b$ baryon)
\bea
\frac{d \Gamma}{dt dz} = \frac{G^2 \alpha_{em}^2 \vel V_{tb} V_{ts}^* \ver^2
m_{\Lambda_b}^2}{384 \pi^5} \sqrt{t^2-r^2} \Bigg\{ A(t) + 
(\vec{s_\Lambda} \cdot \vec{n} ) \frac{\sqrt{t^2-r^2}}{t} B(t) \Bigg\}~,
\eea
where $\vec{s_\Lambda}$ is the spin vector and $\vec{n}$ is the unit vector
along the momentum of the $\Lambda$ baryon, $z=\cos\theta$  and
the functions $A(t)$ and $B(t)$ are expressed as

\bea
A(t) &=& \frac{4 m_b^2 m_{\Lambda_b}\vel C_7^{eff} \ver^2}
{( 1 - 2 t + r^2 )} \Bigg\{\Big[4 ( t - r^2 ) ( 1 - t ) - 
t (1 - 2 t + r^2) \Big] F_1^2 \nnb \\
&+& 2 r 
\Big[4 ( 1 - t )^2 - ( 1 - 2 t + r^2) \Big] 
F_1 F_2 \nnb \\
&+&  \Big[ 8 t ( 1 - t )^2 - 
4 ( t - r^2 ) ( 1 - t ) - t ( 1 - 2 t + r^2 ) \Big] F_2^2 \Bigg\} \nnb \\ 
&+& m_{\Lambda_b}^3 \ga \vel C_9^{eff} \ver^2 + 
\vel C_{10}^{eff} \ver^2 \dr
\Bigg\{\Big[ ( 1 - 2 t + r^2 ) t + 2 ( t - r^2 ) 
( 1 - t ) \Big] F_1^2  \nnb \\
&+& 2 r \Big[ ( 1 - 2 t + r^2 ) + 
2 ( 1 - t )^2 \Big] F_1 F_2 \nnb \\
&+&\Big[( 1 - 2 t + r^2 ) t - 2 ( t - r^2 ) ( 1 - t ) +
4 t ( 1 - t )^2 \Big]F_2^2 \Bigg\} \nnb \\ 
&+& 12 m_b m_{\Lambda_b}^2 \mbox{\rm Re} \ga C_7^{eff} C_9^{*eff}\dr 
\Bigg\{ ( t - r^2 ) F_1^2 +
2 r ( 1 - t ) F_1 F_2 + \Big[ ( t - r^2 ) -
2 t ( 1 - t ) \Big]F_2^2 \Bigg\}~, \nnb \\ \nnb \\
B(t) &=& \frac{4 m_b^2 m_{\Lambda_b}\vel C_7^{eff} \ver^2}
{( 1 - 2 t + r^2 )} \Bigg\{r 
\Big[(1 - 2 t + r^2) - 4 ( 1 - t ) \Big] F_1^2 \nnb \\
&-& 8 ( 1 - t ) r^2 F_1 F_2
+  r\Big[ 8 ( 1 - t )^2 - 
4 ( 1 - t ) - ( 1 - 2 t + r^2 ) \Big] F_2^2 \Bigg\} \nnb \\ 
&-& m_{\Lambda_b}^3 \ga \vel C_9^{eff} \ver^2 + 
\vel C_{10}^{eff} \ver^2 \dr
\Bigg\{ r \Big[( 1 - 2 t + r^2 ) 
+ 2 ( 1 - t )\Big] F_1^2  \nnb \\
&+& 4 \Big[ ( 1 - t ) r^2 \Big] F_1 F_2
-r \Big[( 1 - 2 t + r^2 )  - 2 ( 1 - t ) +
4 ( 1 - t )^2 \Big]F_2^2 \Bigg\} \nnb \\ 
&-& 12 m_b m_{\Lambda_b}^2 \mbox{\rm Re} \ga C_7^{eff} C_9^{*eff}\dr 
\Bigg\{  r F_1^2 + 2 r^2  F_1 F_2 + 
r ( 1- 2t ) F_2^2  \Bigg\} ~,
\eea 
where $r=m_\Lambda/m_{\Lambda_b}$ and $t=E/m_{\Lambda_b}$, respectively.
It should be noted here that $A(t)$ and $B(t)$ were calculated in SM in
\cite{R34} but our results do not coincide with theirs, especially on
$B(t)$. 
Integrating Eq. (16) over $t$, the differential 
decay rate can be rewritten in terms of the polarization variable 
$\alpha$ as
\bea
\frac{d\Gamma}{dz} = \frac{\Gamma_0}{2} \Big\{ 1 +(\vec{s_\Lambda} \cdot
\vec{n} ) \alpha \Big\}~,
\eea
where
\bea
\Gamma_0 = \frac{G^2 \alpha_{em}^2 \vel V_{tb} V_{ts}^* \ver^2
m_{\Lambda_b}^2}{192 \pi^5} \int_{t_{min}}^{t_{max}}
\sqrt{t^2-r^2} A(t) dt~,
\eea 
and $\alpha$ is the asymmetry parameter, whose form is given as
\bea
\alpha=\frac{\displaystyle{\int_{t_{min}}^{t_{max}}
\frac{\sqrt{t^2-r^2}}{t} B(t) dt}}
{\displaystyle{\int_{t_{min}}^{t_{max}}
\sqrt{t^2-r^2}A(t) dt}}~,
\eea
where the integration limits are determined by
\bea
r \le t \le \frac{1}{2} \ga 1+r^2-\frac{4 m_\ell^2}{m_{\Lambda_b}} \dr 
\nnb
\eea
As we noted previously, in model III
a new phase appears. Therefore we would expect larger CP violation
compared to the SM model prediction. The CP violating asymmetry between 
$\Lambda_b \rar \Lambda \ell^+ \ell^-$ and $ \bar \Lambda_b\rar \bar \Lambda
\ell^+ \ell^-$ decays is defined as
\bea
A_{CP} (t) = \frac{\displaystyle{\frac{d \Gamma}{dt} -
\frac{d \bar \Gamma}{dt}}}{\displaystyle{\frac{d \Gamma}{dt}+
\frac{d \bar \Gamma}{dt}}}~.
\eea
The differential widths of the $\Lambda_b \rar \Lambda \ell^+ \ell^-$ and 
$ \bar \Lambda_b\rar \bar \Lambda \ell^+ \ell^-$ decays can easily be 
obtained from Eq. (16) by integrating over $z$. Hence the CP violating
asymmetry takes the following form:
\bea
A_{CP} = - \frac{12 m_b m_{\Lambda_b}^2}{A_1(t)} \mbox{\rm Im}C_9^{eff} 
\mbox{\rm Im}C_7^{eff} 
\Big\{ F_1^2 ( t - r^2) + 2 r ( 1 - t) F_1 F_2 + 
F_2^2 \Big[ ( t - r^2) - 2 t ( 1 - t) \Big]\Big\}~,
\eea
where
\bea
A_1(t) = A(t)\left[ \mbox{\rm Re}(C_7^{eff} C_9^{*eff}) \rar 
\mbox{\rm Re}C_7^{eff} \, \mbox{\rm Re}C_9^{eff}~\right]. \nnb
\eea
In derivation of $A_{CP}$ the following representation of $C_9^{eff}$ and 
$C_7^{eff}$ have been used
\bea
C_9^{eff} &=& \mbox{\rm Re}C_9 + i \mbox{\rm Im}C_9~,\nnb \\
C_7^{eff} &=& \mbox{\rm Re}C_7 + i \mbox{\rm Im}C_7~,
\eea
and following \cite{R34} we assume that the form factors are real.
It should be noted that since Im$C_7^{eff}=0$ in models I, II, and SM,
the CP asymmetry is zero (or suppressed very strongly),
which is one essential difference among the model III and
models I, II and SM.  

\section{Numerical analysis}
In the present work we have considered three different versions, namely
models I, II and III of the 2HDM. For the free parameters
$\lambda_{bb}$ and $\lambda_{tt}$ of model III, we have used the
restrictions coming from $B \rar X_s \gamma$ decay, $B^0$--$\bar B^0$
mixing, $\rho$ parameter and neutron electric--dipole moment \cite{R26},
that yields $\vel \lambda_{bb} \ver = 50$, $\vel \lambda_{tt}\ver \le 0.03$.
Similar analysis restricts the value of $\tan \beta$, which is the free
parameter of model I and model II, to \cite{R35,R36}
\bea
0.7 \le \tan \beta \le 0.6 \ga m_{H^\pm}/1~GeV\dr~, \nnb 
\eea
(the lower limit of the charged Higgs boson mass is obtained 
to be $m_{H^\pm} \ge 200~GeV$ in \cite{R36}).

The values of the main input parameters, which appear in the expressions
for the branching ratio and $A_{CP}$ are: 
$m_b = 4.8~GeV,~m_{\Lambda_b}=5.64~GeV,~m_\Lambda=1.116~GeV,~m_c = 1.4~GeV$. 
The values of the Wilson coefficients are,
$C_1 = - 0.249$, $C_2 = 1.108$, $C_3 = 1.112 \times 10^{-2}$,
$C_4 = -2.569 \times 10^{-2}$, $C_5 = 7.4 \times 10^{-3}$,
$C_6 = - 3.144 \times 10^{-2}$. 
As has already been noted for the form factors that are needed in the
present numerical analysis, we have used the results of the work
\cite{R34}. 

In Fig. (1) we present the dependence of the differential width of the
$\Lambda_b \rar \Lambda \ell^+ \ell^-$ decay on $t$ at $\tan \beta=1.5$ and
at $m_{H^\pm}=250~GeV$ for the models I and II. In this figure we also depict 
the dependence of the same differential decay width at 
$\vel \lambda_{tt} \lambda_{bb} \ver=1.5$ and at the value of the phase
angle $\phi=0$ for the same value of the charged Higgs boson mass. In both
cases the long distance effects are taken into account.

The values of the decay width in three different models of the $2HDM$ for
different choices of the values of the charged Higgs boson mass is listed in
Table 1. It follows from this table that in all three models the charged
Higgs boson contribution to the decay width is negligibly small for the 
values of the 
$\lambda_{tt}$ and $\lambda_{bb}$ (or $\tan\beta$), which lies within the
experimental bounds. Moreover it should be noted that if the long distance
contributions ($J/\psi$ resonances) are neglected, the decay width becomes
two order of magnitude smaller, i.e., $\Gamma_{short dist.} \sim 
8 \times 10^{-19} ~GeV$.

\begin{table}
\begin{center}
\begin{tabular}{|c|c|c|c|}   
\hline
\multicolumn{1}{|c|}{\boss}
&\multicolumn{3}{|c|}{{Decay width {\bf$\Gamma$} $\times 10^{17}$ (in $GeV$)}} \\ 
\hline\hline        
$m_{H^\pm}~(GeV)$&\boss Model I & Model II & Model III \\ \hline
\boss 100  & 6.12 & 6.13 & 6.09 \\ \hline
\boss 250  & 6.10 & 6.10 & 6.08 \\ \hline
\boss 400  & 6.10 & 6.10 & 6.08 \\ \hline
\boss 1000 & 6.09 & 6.09 & 6.08 \\ \hline
\end{tabular}
\vskip 0.3 cm
\caption{}  
\vskip 1 cm 
\end{center}
\end{table} 

The dependence of the asymmetry parameter $\alpha$ on the charged Higgs
boson mass $m_{H^\pm}$ and phase angle $\phi$ in model III 
is presented in Fig. (2) 
when long distance effects are taken into account. We observe from this
figure that the asymmetry parameter $A_{CP}$ increases in magnitude as
the mass of the $m_{H^\pm}$ increases. This is due to the fact that the
decay width increases as $m_{H^\pm}$ increases. It is also observed when for
$0 \le \phi \le \pi$ as modulo $\alpha$ increases and decreases for
$\pi \le \phi \le 2 \pi$. This can be explained by the fact that in the
region $0 \le \phi \le \pi$ ($\pi \le \phi \le 2 \pi$), the charged Higgs
boson contribution to the SM is constructive (destructive).  

For a comparison we present the asymmetry parameter $\alpha$ in SM and 2HDM
with and without long distance contributions, at $m_{H^\pm}=250~GeV$ and
$\phi=0$. 

\bea
\alpha_{2HDM}^{short} &=& -0.48 ~,\nnb \\
\alpha_{2HDM}^{long}  &=& -0.54 ~,\nnb \\
\alpha_{SM}^{short} &=& -0.50 ~,\nnb \\
\alpha_{SM}^{short} &=& -0.54 ~.\nnb
\eea

As we have noted earlier, in model III a new phase appears in 
$\lambda_{tt} \lambda_{bb}$ vertex which is embedded in $C_7^{eff}$ term. As a
result interference of the imaginary parts of $C_7^{eff}$ and $C_9^{eff}$ 
can induce CP violating asymmetry. 

In Fig. (3) we present the dependence of CP asymmetry on $t$ and the phase   
angle $\phi$ in model III. It is observed from this figure that in the 
resonance region $\vel A_{CP} \ver \simeq 4\%$ , and far from resonance 
region $\vel A_{CP} \ver \simeq 1.5\%$. This is a very useful information in
establishing model III, since in models I, II and SM $A_{CP}$ is practically
zero due to the fact that in all these models the Wilson coefficient
$C_7^{eff}$ is real. 

In conclusion, we investigate the $\Lambda_b \rar \Lambda \ell^+ \ell^-$
decay in general 2HDM, in which a new extra phase is present. It is
shown that investigation of the CP asymmetry which is attributed to the 
differential decay width differences, can give unambiguous information about
model III, since in this version CP asymmetry can be quite measurable, while
at the same time CP asymmetry in models I and II are highly suppressed.

\newpage
\section*{Figure captions}
{\bf Fig. 1} The dependence of the differential width of the 
$\Lambda_b \rar \Lambda \ell^+ \ell^-$ decay on $t$, for three different
versions of the 2HDM, at $m_{H^\pm}=250~GeV$. The free parameter of models
I and II is taken $\tan\beta=1.5$ and for model III we choose
$\vel \lambda_{bb}\lambda_{tt} \ver = 1.5$,  $\vel \lambda_{tt} \ver = 0.03$.  
Dotted line represents model I, dash--dotted line represents model II and
solid line represents model III, respectively. \\ \\
{\bf Fig. 2} The dependence of the asymmetry parameter $\alpha$ on
$m_{H^\pm}$ and the phase angle $\phi$ for model III. \\ \\
{\bf Fig. 3} The dependence of CP asymmetry parameter on the dimensionless
parameter $t$ and the phase angle $\phi$ at $m_{H^\pm}=250~GeV$, for model III.  
\newpage
\begin{figure}[H]
\vskip 1.5 cm
    \includegraphics{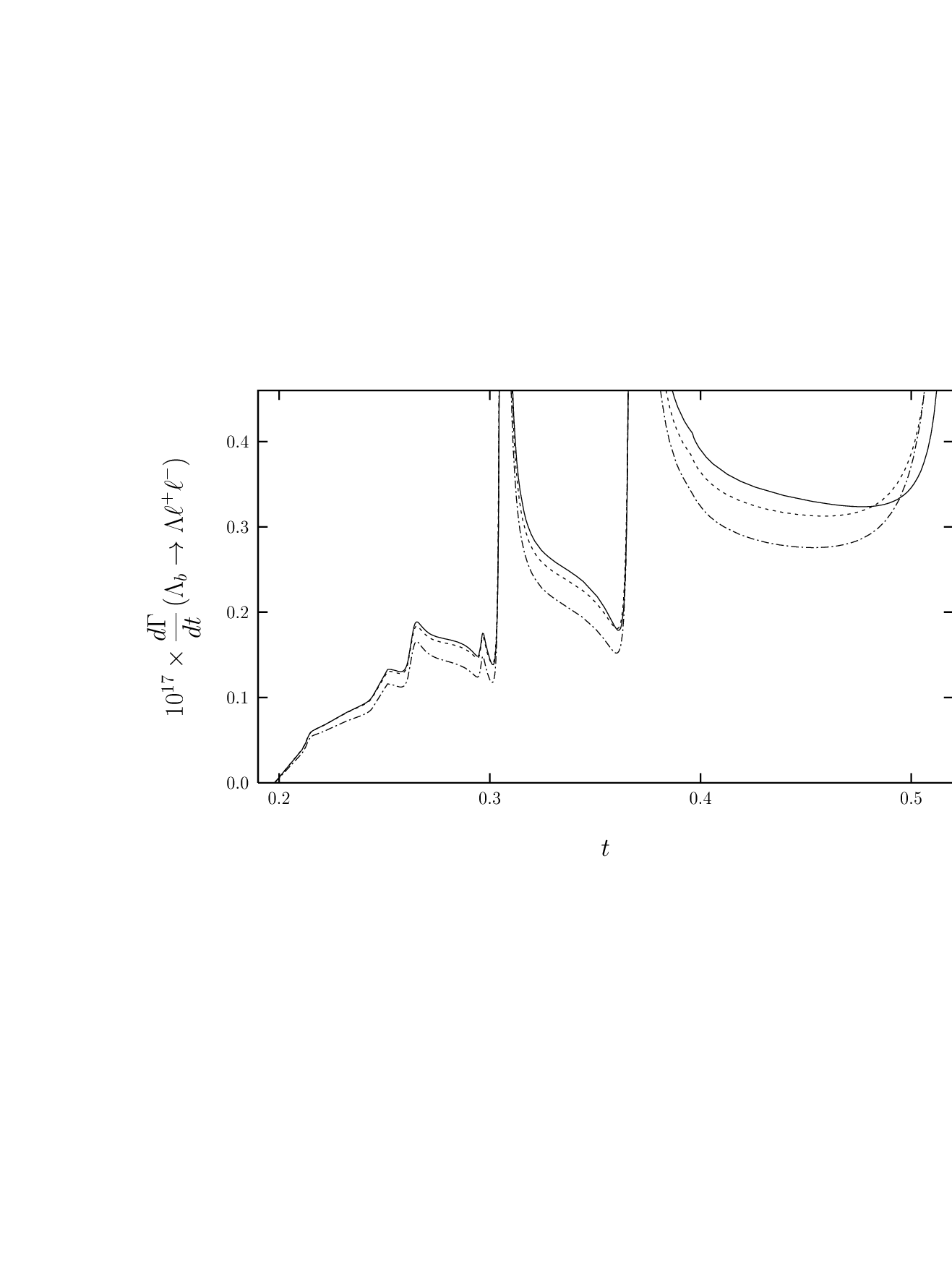}
\vskip 6.7cm   
\caption{}
\end{figure}

\begin{figure}[b]      
\vskip 0 cm
       \includegraphics{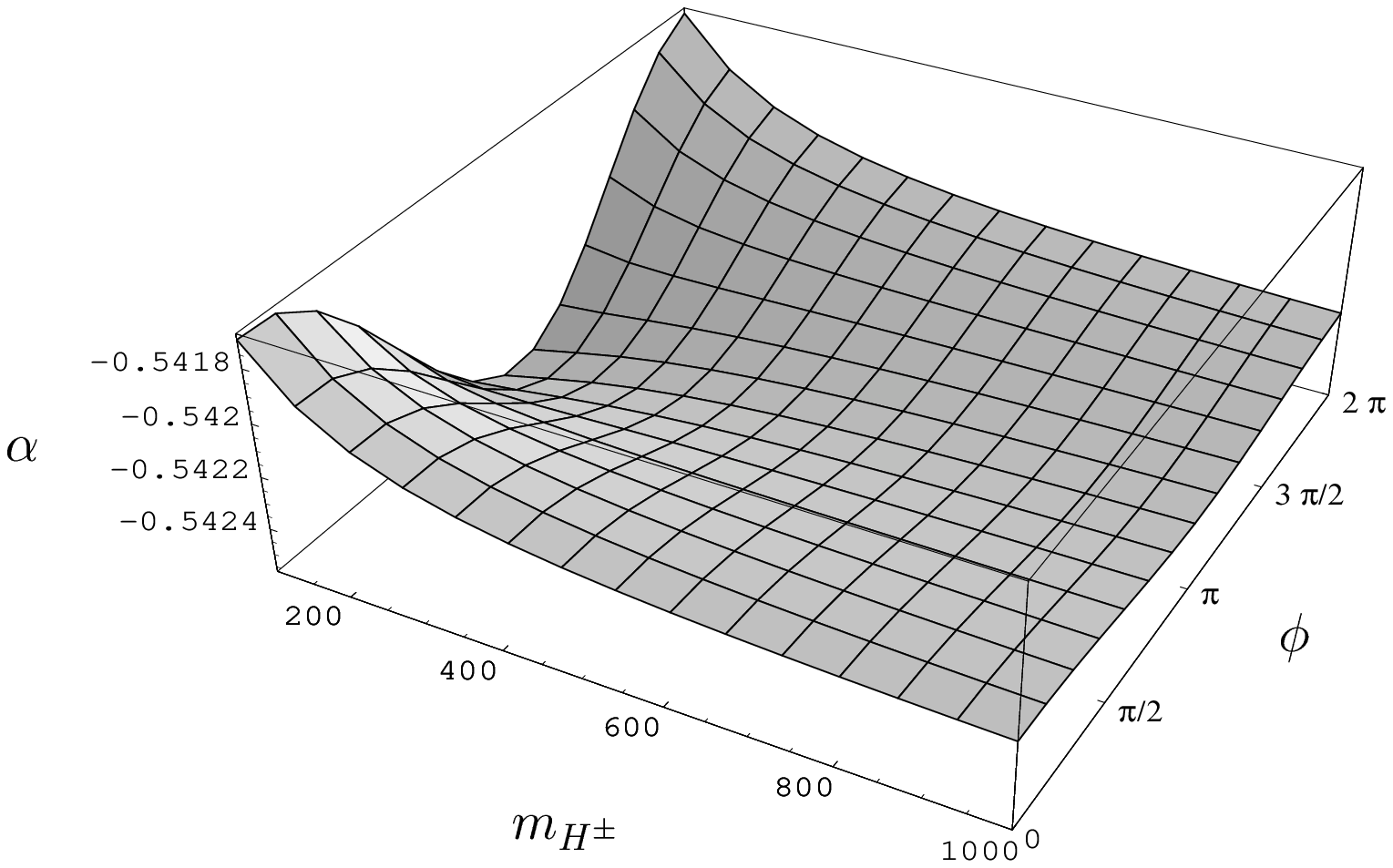}
\vskip 9.5 cm
\caption{ }
\end{figure}

\newpage

\begin{figure}
\vskip 1.5 cm
    \includegraphics{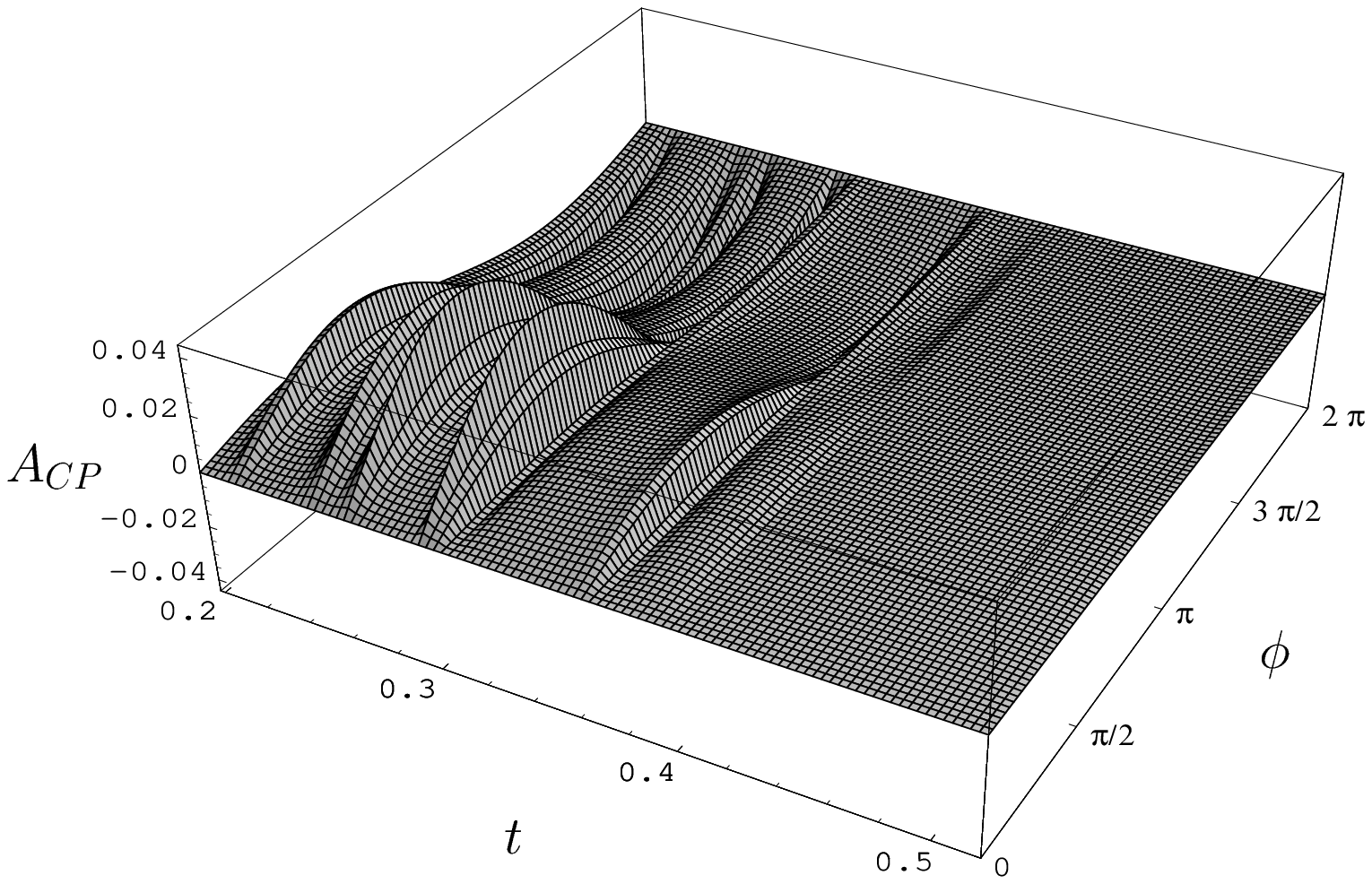}
\vskip 4.7 cm
\caption{}
\end{figure}

\end{document}